\documentclass[sigconf,natbib=false]{acmart}

\AtBeginDocument{%
  }

\copyrightyear{2025}
\acmYear{2025}

\acmConference[Conference acronym 'XX]{Make sure to enter the correct
  conference title from your rights confirmation email}{June 03--05,
  20XX}{Woodstock, NY}

\RequirePackage[datamodel=acmdatamodel, style=acmnumeric]{biblatex}

\usepackage{csquotes} 

\usepackage{stfloats}

\usepackage{tikz}
\usetikzlibrary{arrows.meta}
\usetikzlibrary{intersections}
\usepackage{wheelchart}

\usepackage{hyperref}
\usepackage[english]{babel}
\addto\extrasenglish{

}

\usepackage{comment}

\begin{document}

\title{Perspectives and potential issues in using artificial intelligence for computer science education}

\author{Juho Vepsäläinen}
\email{juho.vepsalainen@aalto.fi}
\orcid{0000-0003-0025-5540}
\affiliation{%
 \institution{Aalto University}
 \city{Espoo}
 \country{Finland}}

\author{Petri Juntunen}
\email{petri.juntunen@aalto.fi}
\orcid{0009-0008-6689-8227}
\affiliation{%
 \institution{Aalto University}
 \city{Espoo}
 \country{Finland}}

\renewcommand{\shortauthors}{Vepsäläinen et al.}

\begin{abstract}
Since its launch in late 2022, ChatGPT has ignited widespread interest in Large Language Models (LLMs) and broader Artificial Intelligence (AI) solutions. As this new wave of AI permeates various sectors of society, we are continually uncovering both the potential and the limitations of existing AI tools.

The need for adjustment is particularly significant in Computer Science Education (CSEd), as LLMs have evolved into core coding tools themselves, blurring the line between programming aids and intelligent systems, and reinforcing CSEd’s role as a nexus of technology and pedagogy.
The findings of our survey indicate that while AI technologies hold potential for enhancing learning experiences, such as through personalized learning paths, intelligent tutoring systems, and automated assessments, there are also emerging concerns.
These include the risk of over-reliance on technology, the potential erosion of fundamental cognitive skills, and the challenge of maintaining equitable access to such innovations.

Recent advancements represent a paradigm shift, transforming not only the content we teach but also the methods by which teaching and learning take place.
Rather than placing the burden of adapting to AI technologies on students, educational institutions must take a proactive role in verifying, integrating, and applying new pedagogical approaches.
Such efforts can help ensure that both educators and learners are equipped with the skills needed to navigate the evolving educational landscape shaped by these technological innovations.

\end{abstract}




\maketitle

\textbf{This version is a preprint for arXiv and it has not been accepted for publication yet! Send us feedback to help us improve the paper for publication in a suitable venue.}

\section{Introduction}
\label{sec:introduction}
Computer science has historically served as a fertile ground for educational innovation, given that many technological advancements originate within the discipline itself \cite{tucker1996strategic}.
The most recent wave of innovation in Computer Science Education (CSEd) centers on the integration of Artificial Intelligence (AI) into teaching, and its far-reaching implications for both instructional practices and learning experiences.

ChatGPT, launched in late 2022, renewed interest specifically in generative AI \cite{fui2023generative}, and this development has opened up a wide array of possibilities for educators, instructional designers, and researchers to explore novel pedagogical approaches. As AI technologies become increasingly integrated into educational systems, they present promising opportunities to augment instructional practices, personalize learning experiences, and optimize the use of institutional resources, potentially facilitating the development of more effective, equitable, and sustainable learning environments. 
Despite their promise, AI technologies also raise concerns, particularly as best practices for their effective use in education are still evolving. Ensuring that these tools genuinely enhance learning outcomes, especially within the context of increasingly constrained teaching resources worldwide, requires careful, evidence-informed implementation.

\subsection{The current state of AI technologies}

The development of AI has historically been characterized by alternating periods of enthusiasm and decline, often referred to as "AI Summers" and "AI Winters", reflecting intervals of heightened interest and subsequent skepticism regarding the technology's potential \cite{kautz2022third}.
According to Kautz \cite{kautz2022third}, the latest "AI Summer" began around 2012 or 2016, depending on one's perspective, and gained significant momentum with the launch of ChatGPT in November 2022.
Since then, the field has attracted substantial public attention, with the capabilities and applications of AI tools expanding considerably.
Today's reasoning models are no longer limited to text; they have evolved to become multimodal, capable of processing various media types, including images, structured markup, and code.
This versatility makes them valuable tools for pattern-matching and problem-solving tasks across various domains.

\begin{figure}[ht]
  \centering
  \includegraphics[width=\linewidth]{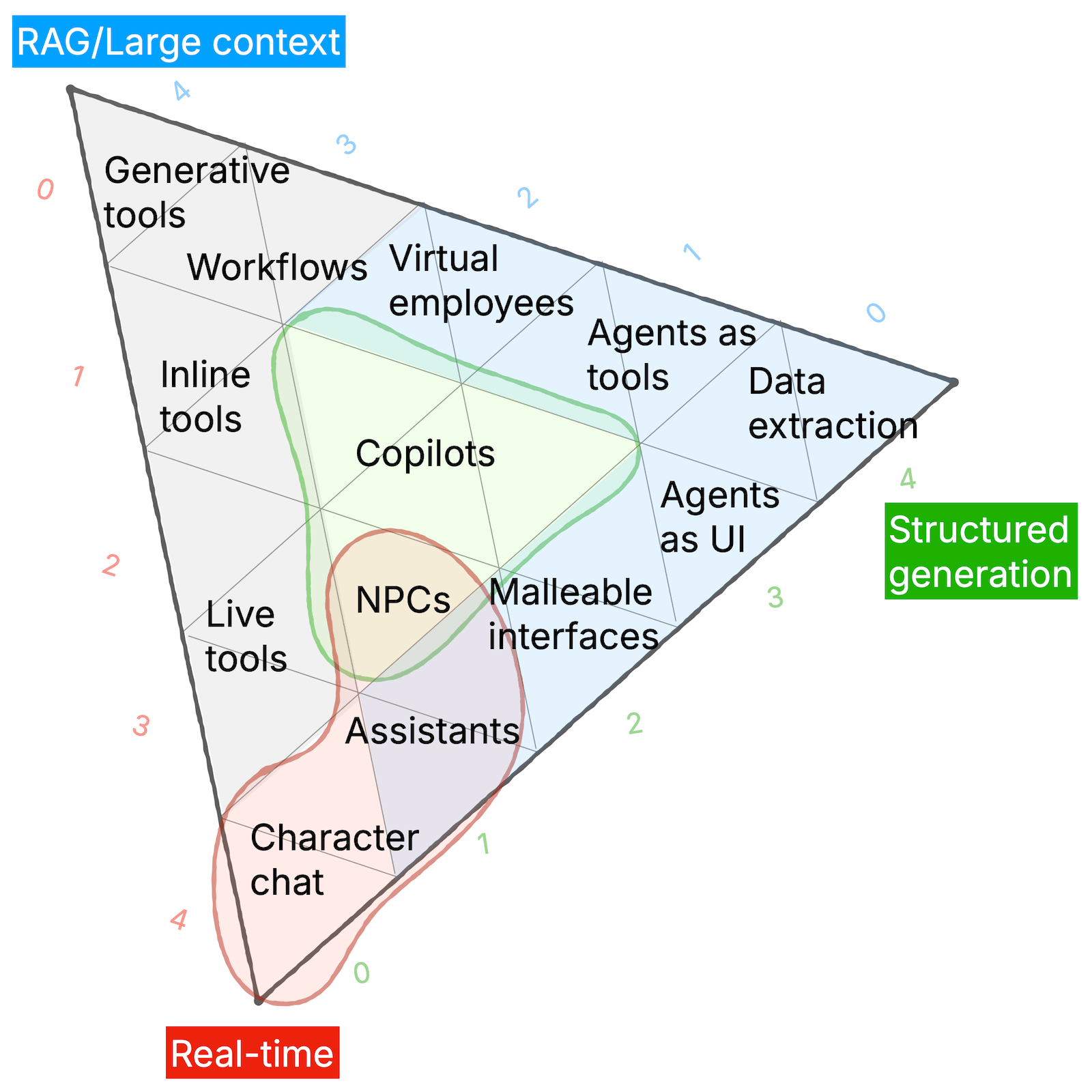}
  \caption{Webb \cite{webb2024} classifies AI tools based on dimensions of RAG/Large context, real-time, and structured generation.}
  \label{fig:ai-tools}
  \Description{The triangle contains 25 subdivisions as smaller triangles that capture types of tools. Starting from RAG/Large context corner, the types are as follows: generative tools, inline tools, workflows, virtual employees, live tools, copilots in the middle, agents as tools, character chat in the real-time corner, assistants, malleable interfaces, agents as UI, and data extraction in the structured generation corner.}
\end{figure}

Webb \cite{webb2024} classifies AI tools within the dimensions of RAG/Large context, real-time, and structured generation as illustrated in Figure \ref{fig:ai-tools}.
Although not exhaustive, this framework provides a lens for contextualizing the landscape of AI technologies in education.
Webb’s classification highlights the diversity of AI applications, ranging from chatbots and copilots to assistants, agents, and virtual employees, each representing distinct technical capabilities and potential use cases.

\subsection{Software industry’s transition toward AI}

Recent advancements in AI technologies are reshaping the software industry in profound ways, as discussed by Kokol \cite{kokol2024use}.
In his analysis, Kokol showed that AI has changed the way engineers develop software, and more specifically, it has changed the type of work engineers perform, as AI can take over some of the tasks.
As the field undergoes a radical transformation, educational models must develop in parallel, with required skill sets continuously evolving to meet the emerging demands.
This ongoing transition is reflected in the ACM’s CS2023 model curriculum, where AI is more prominently featured than in any previous iteration \cite{servin2024cs2023}.

\subsection{Why investigate the use of AI in CSEd?}

The integration of artificial intelligence within CSEd is an emerging area of research that has attracted growing attention in recent years.
Even without formal changes to educational systems, various stakeholders, including students, educators, and institutions, are likely to adopt AI tools due to their convenience and powerful capabilities.
Consequently, it is essential to comprehend both the potential benefits and inherent challenges associated with AI utilization in CSEd.
This need is underscored by the rapid evolution of the field and a growing body of scholarly literature, including systematic reviews, that aim to map and critically assess this emerging area \cite{vierhauser2024towards,raihan2025large,prather2025beyond,al2024impact,wang2025effect,memarian2023chatgpt,nwoyibe2025deploying,cai2024exploring,kumar2025ThePI,gaitantzi2025role,chugh2025promise,tang2024ImplicationsOA,deckker2025AIPoweredPL,zhu2025systematic,chang2024systematic}.

Notably, Beale \cite{beale2025computer} has covered the usage of AI in CSEd from a practical angle and considered its impact on teaching practices and policies.
Prather et al. \cite{prather2025beyond} note that educators are already adapting their teaching approaches by refocusing their programming curricula and introducing new topics, such as prompt engineering, establishing policies for when to use AI, incorporating AI into lectures, and teaching students how to use AI effectively highlighting the importance of researching the space as technology is actively being adopted.
Given the way integration of AI is going, it is not a question of when but a question of how, and that is where research can help to calibrate the direction.

\subsection{Research questions}

As a topic, CSEd in AI is a complex one to tackle since the field is moving fast at the moment, and new research is becoming available constantly.
To frame this article, our main purpose is to capture a snapshot of the current perspectives while considering potential issues, since it is those that we have to be aware of and tackle during the adoption of technologies.
Based on this background, we have set our research questions for this survey as follows:


\begin{enumerate}
    \item What are the prevalent perspectives surrounding the adoption of AI within CSEd?
    \item What are the issues potentially hindering the adoption of AI in CSEd?
\end{enumerate}

\subsection{Research approach and contribution}

We used topical searches against \href{https://scholar.google.com/}{Google Scholar}, \href{https://www.semanticscholar.org/}{Semantic Scholar}, and \href{https://www.researchgate.net/}{ResearchGate} services to discover articles to read.
Occasionally, reading articles led to more timely content to consider, and given the vast amount of sources available in the space, it is more than likely that we missed good ones.
The main purpose of our research was to capture the main themes around the topic and structure them in a narrative, allowing discussion in a focused manner, although the topic itself is broad and warrants more specific consideration.
The primary contribution of this article is a collection of perspectives and arguments you can use to consider when adopting AI technologies in CSEd.
Furthermore, by establishing the main prevalent perspectives and issues, we make it easier for subsequent researchers to develop the discovered themes further and perhaps add new categories to consider.





\subsection{Structure of the paper}

To address the first research question and incorporate diverse viewpoints, \autoref{sec:perspectives} presents existing perspectives on the adoption of AI in CSEd, and \autoref{sec:issues} reviews AI adoption-related issues discovered from the literature to cover the second research question.
\autoref{sec:discussion} then explores the implications of these findings while considering the future, and \autoref{sec:conclusion} summarizes the key observations while outlining potential directions for future research.

\section{Perspectives on the adoption of AI in CSEd}
\label{sec:perspectives}
The introduction of the printing press in the 15th century led to the proliferation of printed media and arguably revolutionized the societies benefiting from the new invention \cite{lee22}.
Although the impact of printing can be seen beneficial in retrospect, printing had its opponents, such as Filippo de Strata, who argued that the economics of printing allowed poor quality material to circulate in public \cite{sabev2009rich}.
Similar arguments can be constructed against the usage of AI, and in this section, we will first capture current AI-critical and AI-positive views before considering what AI usage looks like from the perspective of educators and students.

\subsection{AI-critical view}

\paragraph{The introduction of AI into learning environments presents several notable risks that warrant careful consideration}
Wang et al. \cite{wang2023exploring} highlight the inherent risks associated with AI tools, particularly concerning misuse and disparities in learning. Students who are unable to effectively utilize these tools may find themselves at a disadvantage compared to those who can.
Wang et al. examined how CS educators might make AI tools more approachable for students and found that many preferred real-time, in-class activities to engage and assess student learning, rather than relying on generative AI tools.
Wang et al. identified potential in the use of generative AI tools for automatic problem modification. However, respondents expressed concerns that these tools would necessitate manual verification, subsequently increasing the potential workload.
Wang et al. also identified challenges related to the trustworthiness of generative AI tools, emphasizing the need to prioritize knowledge acquisition over blind reliance. They note that the effectiveness of these tools often depends on precise prompting, as their output can be inconsistent and highly sensitive to how queries are formulated.

\paragraph{AI should be framed as collective intelligence}
MacArthur \cite{macarthur2025large} highlights that AI comes with significant risks due to multiple factors including automation bias\footnote{Automation bias means putting too much trust in machine-generated suggestions \cite{macarthur2025large}.}, experience bias\footnote{Experienced people are in a better spot to judge AI output therefore making it difficult for beginners to refine their prompts \cite{macarthur2025large}.}, and misleading terminology.
MacArthur argues that AI technologies are best conceptualized as a form of collective intelligence, reflecting the dynamic interplay between human cognition and technological systems. Rather than viewing AI as a replacement for human thought, it should be emphasized in its role in augmenting human capabilities. In this context it the relevance of the humanities, particularly the skills of critical reading, writing, and editing, should be emphasised as essential for interpreting, questioning, and shaping the outputs of AI systems. These skills, she argues, are not diminished by the rise of AI but are more crucial than ever for fostering thoughtful engagement, ethical reflection, and informed participation in increasingly AI-mediated environments.


\paragraph{Pedagogic reductionism puts focus on technology over pedagogy}
Zeide \cite{yuan2024does,zeide2019artificial} points out that reliance on platforms developed by for-profit corporations carries the risk of pedagogical reductionism, where emphasis may be placed on technology rather than educational principles. This risk is pertinent to AI platforms, many of which are controlled by commercial entities. These companies might prioritize increasing their market share by tightly integrating with university programs, potentially aligning with their interests over pedagogical goals.

\paragraph{Tools affect our learning}
Kosmyna et al. \cite{kosmyna2025your} showed that how we employ AI tools can significantly influence our brain activity and learning outcomes. Their early findings resonate with Woodard's \cite{woodard2018mixed} observations on the effects of calculator usage in educational settings. These studies underscore the critical importance of utilizing tools in a way that enhances learning rather than detracts from it.

\paragraph{AI should not be adopted uncritically}
Open letters \cite{openletter2025,dusseau2025} against the uncritical adoption of AI in academia emphasize the importance of carefully evaluating how AI is integrated by universities.
Specifically, the letters advocate for critical evaluation of the incorporation of AI into university software systems, prohibiting its use in classrooms, countering the normalization of AI hype, strengthening academic freedom, and maintaining a critical perspective on AI practices.
The open letters serve as a valuable reminder of the core mission of universities and their primary role as educators.

\subsection{AI-positive view}

\paragraph{AI enables new types of teaching approaches}
Mollick \cite{ethan23} has embraced an AI-friendly approach in his classes, encouraging the use of AI tools with the understanding that students will likely utilize these technologies regardless.
As part of his instruction, Mollick invites students to use ChatGPT to draft essays through prompting, demonstrating the importance of crafting precise prompts and enhancing them through prompt engineering.
Furthermore, Mollick and Mollick \cite{mollick2023using} found that specific teaching strategies, such as low-stakes testing or continuous assessments, have become feasible through the use of AI tools. Consequently, AI can be considered a force multiplier in teaching when utilized effectively by a knowledgeable educator who is adept at navigating the challenges (such as hallucinations) associated with this emerging technology.
Mollick's approach exemplifies an AI-positive perspective, where the advantages of AI are perceived to outweigh its drawbacks. Recognizing that students will inevitably use AI, Mollick advocates for guiding them in its appropriate use, rather than allowing them to learn the tools in an ad hoc manner that may produce suboptimal results.

\paragraph{AI as a potential solution to the two-sigma problem}
Bloom's two-sigma problem, identified in the 1980s, highlights the challenge of delivering personalized feedback at scale. While the advantages of individualized tutoring are well-established and documented, they have traditionally been difficult to implement cost-effectively. However, the emergence of generative AI technologies now offers promising solutions to address this issue \cite{de2025chalkboards,mollick2024co,hartley2024artificial}.
De Simone et al. \cite{de2025chalkboards} experimented in Nigeria, under challenging conditions, aiming to solve the two-sigma problem. Their research provided initial evidence that the problem is indeed solvable. By assessing the advantages of personalized learning facilitated by generative AI, De Simone et al. identified significant benefits for the students involved in the trial. Their intervention emerged as a potentially cost-effective method for enhancing educational standards.
Furthermore, to underscore this point, Yusof \cite{yusof2025chatgpt} found that personalized feedback facilitated by AI significantly enhanced the long-term retention of topics when comparing a group of students utilizing AI with a group that did not.

\paragraph{AI and the Zone of Proximal Development}
Vygotsky's Zone of Proximal Development (ZPD) \cite{vygotsky1987zone} describes the range within which learners can perform tasks independently or with assistance.
The concept of ZPD complements the two-sigma problem, as AI has the potential to empower students to accomplish more autonomously, thereby enhancing both learning outcomes and motivation.
Cai et al. \cite{cai2024exploring} conducted a systematic literature review on this topic and concluded that AI tools align effectively with the principles of the ZPD. These tools appear capable of adapting to students' needs and fostering collaborative interactions. 

\paragraph{AI may speed up learning while improving learning outcomes}
In their investigation of a mathematics classroom integrating AI, Xu \cite{xu2024ai} documented a range of positive outcomes, including enhanced learning achievement, increased motivation and engagement, more timely and personalized feedback, heightened self-efficacy and learner confidence, more targeted instructional support based on individual needs, and improved transfer of knowledge.
Overall, the study reconceptualized the mathematics learning environment by utilizing AI as a learning platform personalized per student, resulting in potential pedagogical benefits.
The early results of Kestin et al. \cite{kestin2024ai} corroborate Xu's findings, demonstrating that students receiving personalized tutoring via AI learned at twice the rate of those in the control group.

\paragraph{AI-driven learning analytics: enabling new educational interventions}
Demartini et al. \cite{demartini2024artificial} emphasize the potential of AI-driven learning analytics to provide educators with deeper insights into student performance, potentially allowing them to anticipate challenges such as dropouts.
However, the implementation of learning analytics introduces significant ethical and privacy challenges, requiring careful planning, cross-institutional coordination, and legislative initiatives to address these issues and enable effective use of AI in both direct teaching and student performance analysis. Demartini et al. are actively investigating these issues in their Data2Learn@Edu project.


\subsection{Educator perspectives on AI usage}

As shown by Wang et al. \cite{wang2023exploring}, educators hold conflicting views on the use of AI in education.
While the proliferation of AI tools has raised concerns about academic integrity, Wang et al. highlight that educators recognize the potential value of these tools, provided they are used appropriately.
To mitigate these concerns, educators in Wang et al.'s study expressed a preference for incorporating more in-class activities, with one even suggesting the adoption of flipped learning\footnote{Flipped learning is a teaching approach where class time is used for engaging students in collaborative, hands-on activities while direct instruction occurs outside of classes \cite{karabulut2018systematic}.} over traditional teaching methods.
Educators in Wang et al.'s study identified faulty mental models arising from AI tools and their potential for generating hallucinations as a major concern, and this is in line with the observations of Ben-Ari \cite{ben2001constructivism} that bricolage, the act of constructing models from individual pieces, can result in incomplete learning in computer science.
Additionally, educators in the study expressed concerns about the use of AI in introductory courses, noting that it could encourage shallow learning. This concern arises from the fact that current AI tools can easily solve entry-level exercises, a challenge that may be less pronounced in advanced courses with more complex problems.

Zastudil et al. \cite{zastudil2023generative} discovered that educators believe AI tools can aid students in comprehending code and computing concepts, finding inspiration, brainstorming, and receiving feedback on their ideas.
Furthermore, the affordability and accessibility of AI tools enable students to access high-quality learning resources that were previously out of reach.
This last point goes back to De Simone et al. \cite{de2025chalkboards} observation about how AI tooling can facilitate rapid feedback loops for students, potentially improving learning outcomes cost-effectively.
However, as a counterpoint, in Zastudil et al.'s study, educators expressed concerns about trustworthiness, over-reliance, and academic integrity, underscoring the need to develop teaching approaches that discourage misconduct and suboptimal learning processes.

Prather et al. \cite{prather2025beyond} further underscore concerns related to academic integrity, highlighting how the widespread availability of AI tools has significantly lowered the barriers to academic misconduct.
The ease of access to AI-driven solutions can inadvertently encourage students to bypass genuine learning efforts, thus undermining the educational process as a whole.
These concerns emphasize the need for institutions to develop comprehensive strategies and policies that promote the ethical use of AI while upholding rigorous academic standards.

Selwyn et al. \cite{selwyn2025} found out through 57 interviews of teachers from Sweden and Australia that the usage of GenAI may yield more work for educators as they have to consider how to introduce AI to their classroom while working around its limitations to make it useful.
Specifically, Selwyn et al. note that AI tools cannot replace human judgment.

\subsection{Student perspectives on AI usage}

To gauge student opinions on AI usage, Kumar and Raman \cite{kumar2022student} surveyed to identify which educational processes are most suitable for AI integration from the students' perspective. The survey revealed that students generally support the application of AI in teaching and administrative tasks, aligning with existing literature. However, students justifiably expressed apprehension about using AI for admissions, examinations, and placements due to concerns over accuracy and reliability.

Zastudil et al. \cite{zastudil2023generative} discovered through their survey that AI tools significantly facilitated students' coding processes and access to learning materials. These tools allowed students to bypass mundane tasks, often perceived as repetitive and trivial, and enabled them to concentrate on higher-level concepts while benefiting from diverse perspectives and additional assistance. However, both educators and students raised concerns about potential over-reliance on AI tools, questioning their trustworthiness and impact on academic integrity. While students acknowledged that opportunities for academic misconduct existed before the emergence of AI tools, they noted that these technologies have lowered the threshold for such behavior by making it more accessible and less consequential, particularly due to the widespread availability of free AI tools.

Sousa and Cardoso \cite{sousa2025use} conducted a survey on the use of GenAI among Portuguese higher education students and found that the perceived usefulness and ease of use of GenAI tools are defining factors influencing their adoption. However, students also expressed concerns regarding academic misconduct and the reliability of GenAI systems, recognizing the necessity for verifying the information these tools generate. 
While such concerns indicate a level of critical awareness, it remains uncertain to what extent this awareness translates into actual behavior. As noted in theories of behavior change, intention alone does not guarantee action; perceived behavioral control and contextual factors often mediate the relationship between attitudes and behavior. Similarly, from a behaviorist perspective, unless reinforced by clear institutional consequences or incentives, concerns may not result in meaningful shifts in student conduct. Therefore, although students may cognitively recognize the ethical and practical limitations of GenAI, this does not necessarily lead to responsible or restrained use in practice, particularly in the absence of structured guidance, accountability mechanisms, and supportive learning environments.

\subsection{Summary}

Perspectives on the adoption of AI in CSEd tend to vary from critical to positive, and likely many practitioners exhibit views from both extremes, as it is healthy to be both skeptical of development while preparing for the best.
Educators and students seem to have different perspectives of AI shaped by their vantage since especially from the student point of view, aspects such as the need for privacy stand out.
For educators, AI seems to provide new ways to help their students grasp computing-related concepts more easily while allowing students to gain faster feedback, although there are concerns about the trustworthiness of available tools and academic integrity due to easy access.

\section{AI-related issues and their remediations}
\label{sec:issues}
Regardless of one's perspective, computer science educators face a myriad of challenges that are likely to persist and evolve over the coming years.
This section examines critically a spectrum of emerging issues poised to influence the trajectory of computer science education. Drawing on technological trends, pedagogical research, and scenario-based foresight, we interrogate the assumptions underlying current practices and highlight the complex interplay between innovation and institutional inertia.
These dynamics present are not only novel opportunities for pedagogical advancement but also significant challenges that demand critical reflection and strategic response from educators and policymakers alike.
Specifically, we focus on the themes of inequality, technological challenges, biases, and learning challenges as AI-related issues can be grouped within these rough categories.

\subsection{Inequality}

A significant concern within the realm of education is the potential exacerbation of inequalities among students, particularly as they engage with artificial intelligence technologies.
As many have observed \cite{wang2023exploring,prather2025beyond,ignas2025ExploringTE,onet2025TheUO}, disparities in access and effectiveness may arise when students with varying levels of proficiency and diverse learning habits interact with AI tools.
Research indicates that students who are already performing well academically are positioned to derive the greatest benefits from these technologies.
These students tend to leverage AI tools to enhance their learning experiences and expand their intellectual capabilities further.
In contrast, students who face challenges in academics may struggle to exploit these tools effectively due to a variety of barriers.
Ignas \cite{ignas2025ExploringTE} raises critical considerations regarding the limitations of freely available AI tools, which may not provide comprehensive functionality compared to their subscription-based counterparts.
Also, the financial burden associated with premium subscriptions can be prohibitive for economically disadvantaged students, further widening the access gap.
Additionally, the effective use of AI tools often requires reliable internet connectivity, a resource that is not universally available, particularly in under-resourced communities.

Modi and Garg \cite{modi2025TheRO} underscore the concern that access to AI tools can potentially exacerbate existing educational disparities, particularly between universities with ample funding and those that are under-resourced, as well as between urban and rural educational environments.
However, De Simone et al. \cite{de2025chalkboards} provide evidence suggesting that targeted interventions in underprivileged locations can indeed be effective, given that sufficient resources are allocated. Their research demonstrates that strategic measures, such as the deployment of technology infrastructure and the provision of comprehensive training programs, can mitigate many of the adverse effects of resource constraints and enable students in these areas to access and benefit from AI tools.

\subsection{Technological challenges}

\paragraph{Focusing on technology over pedagogy}
Focusing excessively on technology at the expense of pedagogy, as noted by many \cite{zeide2019artificial,govindaraja2024enhancing,zastudil2023generative,singh2025transformative,kumar2022student,grigoryan2025ARTIFICIALII,selwyn2025}, may result in shallow learning as noted by various papers \cite{wang2023exploring,kosmyna2025your,nguyen2024artificial,beale2025computer,perifanou2025collaborative}.
While students may acquire superproficiency in utilizing technological tools, they risk failing to achieve a deep understanding of the underlying content.
This approach could also lead to an over-dependence on these technologies, a concern echoed by several researchers \cite{zastudil2023generative,nguyen2024artificial,beale2025computer,mubashir2025HUMANAA,lyandaIntegratingAI}.

Oneţ \cite{onet2025TheUO}, Solomon and Devi \cite{solomon2025StudentPA}, and Grigoryan et al. \cite{grigoryan2025ARTIFICIALII} raise concerns that an increasing dependence on AI-driven tools in educational settings may marginalize the role of human interaction, an essential component of the learning process. This shift risks reducing education to a transactional process of information delivery, thereby eroding its deeper purpose: the cultivation of critical thinking, dialogic engagement, and holistic intellectual development. 
As a potential solution, Wang et al. \cite{wang2023exploring} and Beale \cite{beale2025computer} propose using alternate teaching methods, such as flipped learning.
Bastani et al. \cite{bastani2025generative} emphasize the necessity of implementing guardrails when utilizing generative AI models, such as GPT, as educational tutors. This perspective aligns with Liu et al. \cite{liu2025}, who advocate for a careful and thoughtful introduction of AI technologies to students. The integration of AI into educational environments must be approached with strategic consideration to ensure that these tools enhance learning rather than detract from it.
Here, how AI is presented to students, the "packaging", is critical. Many AI tools currently available are not inherently designed with educational purposes in mind; thus, educators are tasked with the critical responsibility of selecting and integrating these technologies into their teaching practices. This includes assessing their suitability for educational use and rigorously measuring their impact on student learning outcomes.

\paragraph{Mismatch between AI tooling and educational purpose}
AI tools like ChatGPT are inherently designed for answering questions rather than serving as mentors, a distinction that presents pedagogical challenges, as highlighted by Liu \cite{liu2025}.
To address this limitation, Wang et al. \cite{wang2023exploring}, suggest refining AI tools to be more tailored and specialized for educational contexts, while also encouraging learning through real-time, interactive in-class activities. This hands-on approach aims to redefine the traditional role of classroom instruction by transforming classes into active learning environments.
MacNeil et al. \cite{macneil2025fostering} addressed a key challenge in computer science education by embedding an interactive quiz within a simulated IDE environment. This intervention was designed to prompt students to critically evaluate code suggestions generated by the editor. The results were promising, indicating that the approach effectively enhanced the students’ ability to discern between high- and low-quality suggestions.

\paragraph{Instruction dilution}
Liu et al. \cite{liu2025} have been experimenting with AI in CSEd for years using customized chatbots, most notably \href{https://cs50.ai/}{CS50 Duck}.
By November 2024, approximately 211,000 students had incorporated the duck into their learning as a part of an online course offering. While the primary objective of the CS50 Duck was to provide guidance rather than direct solutions to problems, the sheer scale of the system sometimes resulted in AI-generated responses that did not fully align with specific teaching objectives, and this phenomenon has been termed "instruction dilution."
Moreover, Liu et al. acknowledge the absence of a robust mechanism for evaluating the performance and efficacy of their AI setup, which complicates the assessment of whether the system meets the educational expectations of instructors and students alike. This gap in evaluation further contributes to the risk of instructional misalignment.

To address these challenges, Liu et al. \cite{liu2025} have proposed several remedial strategies. First, the integration of a human-in-the-loop approach allows for continuous supervision and adjustment of AI interactions to ensure they conform to teaching objectives. Second, employing few-shot prompting enhances the quality of AI responses by providing targeted examples that guide the system's outputs more effectively. Lastly, fine-tuning the model involves adjusting the AI's parameters to better fit specific educational contexts, thus reducing the risk of instruction dilution.
Further support for fine-tuning models in educational contexts is provided by Solano et al. \cite{solano2025narrowing}, whose findings imply that a small, fine-tuned model can achieve pedagogical performance comparable to that of a significantly larger counterpart. This suggests that developing domain-specific models is a viable and potentially efficient approach for educational applications.

\paragraph{Trustworthiness}
The trustworthiness of AI tools has emerged as a critical concern in educational settings, particularly when their outputs are unreliable or opaque as shown by clear interest in the topic by many \cite{cumming2024towards,li2023trustworthy,aler2024teach,wang2023exploring,mollick2023using,zastudil2023generative,zeide2019artificial,znamenskiy2025integrating,ignas2025ExploringTE,kumar2025ImpactOA,perifanou2025collaborative,omar2024redefining,selwyn2025}. When students and educators cannot confidently rely on AI-generated results, the educational value of such tools is called into question. To address this challenge, Wang et al. \cite{wang2023exploring} advocate for prioritizing knowledge acquisition over blind acceptance of AI suggestions, emphasizing the importance of fostering students’ critical engagement with AI outputs.
As described by Afroogh et al. \cite{afroogh2024TrustIA}, the concept of trustworthy AI serves as a unifying theme that invites critical examination of trust across various contexts in which AI systems are deployed.
A persistent issue with current-generation AI tools is their tendency to produce hallucinations; outputs that appear plausible but are factually incorrect. This limitation underscores the necessity of verification and poses significant challenges in educational settings, where accuracy and reliability are foundational. In response, domain-specific solutions, such as the system developed by Liu et al. \cite{liu2025}, have the potential to not only mitigate the risks associated with hallucinations but also enable AI systems to align more closely with pedagogical goals.
Another issue pointed out by Solomon and Devi \cite{solomon2025StudentPA} and Oneţ \cite{onet2025TheUO} has to do with the lack of transparency in the tools, as it is difficult to understand how they came up with their results.


\paragraph{Ethics, security, and privacy}
AI ethics is a large theme by itself as shown by a large amount of literature available \cite{tian2024ai,openletter2025,govindaraja2024enhancing,zeide2019artificial,nguyen2024artificial,beale2025computer,singh2025transformative,sun2025theoretical,modi2025TheRO,kumar2025ImpactOA,lyandaIntegratingAI,ahma2025HarnessingAI,grigoryan2025ARTIFICIALII,deckker2025AIPoweredPL,stoic2025TheEO}.
Security and privacy by extension are large issues by themselves, and they have been covered by many papers \cite{tian2024ai,govindaraja2024enhancing,zeide2019artificial,denny2024computing,nguyen2024artificial,modi2025TheRO,onet2025TheUO,solomon2025StudentPA,perifanou2025collaborative,ahma2025HarnessingAI,omar2024redefining,deckker2025AIPoweredPL,stoic2025TheEO} as they are common concerns.

\subsection{Biases}

Utilizing AI technologies inherently involves a spectrum of biases related to their design, technological framework, and application. Among these, the most critical biases to acknowledge are automation bias, experience bias, and social/algorithmic bias.

\paragraph{Automation bias}
Automation bias, as detailed by MacArthur and Peterson \cite{macarthur2025large,peterson2025AIGI}, refers to the undue trust placed in machine-generated outputs, often at the expense of human judgment. This bias presents considerable challenges, as individuals may accept AI-generated information without sufficient scrutiny, potentially leading to erroneous conclusions and decisions. 
To combat automation bias, MacArthur \cite{macarthur2025large} advocates for emphasizing the development of critical skills such as reading, writing, and editing alongside technical proficiency, and this is in line with the warning of Omar and Thomas \cite{omar2024redefining} against erosion of soft skills through over-reliance on AI.
This approach aims to empower individuals to critically assess AI-generated content, fostering a more discerning use of the output.

\paragraph{Experience bias}
Experience bias \cite{macarthur2025large} refers to the advantage held by individuals with prior experience when interacting with prompt-based AI tools, as effective use often depends on the ability to articulate requests in precise and nuanced ways. This dynamic can marginalize less experienced users and exacerbate existing disparities in learning outcomes. To address this issue, MacArthur \cite{macarthur2025large} advocates for the development of strong domain-specific expertise, which can empower users to engage more effectively with AI systems and mitigate the effects of this bias.

\paragraph{Social/algorithmic bias}
Social and algorithmic biases, as discussed by many \cite{tian2024ai,sharma2021social,denny2024computing,modi2025TheRO,onet2025TheUO,lyandaIntegratingAI,solomon2025StudentPA,ahma2025HarnessingAI,stoic2025TheEO}, can emerge during the development of AI tools due to the encoding of certain social ideas within their training data. These biases can inadvertently reflect and perpetuate societal inequities, manifesting in AI outputs.
To address the issue of social bias, Sharma et al. \cite{sharma2021social} emphasize the importance of recognizing and accounting for these biases during the development phase of AI systems. This awareness is crucial for designing AI tools that are more equitable and unbiased. As Zeide \cite{zeide2019artificial} exemplifies, specific design strategies can be implemented to mitigate biases related to race or gender within training datasets.

\subsection{Learning challenges}

\paragraph{Academic integrity}
Academic integrity is widely recognized as a pressing concern across the literature \cite{wang2023exploring,zastudil2023generative,prather2025beyond,humble2024cheaters,nguyen2024artificial,ignas2025ExploringTE,onet2025TheUO,omar2024redefining}.
The increasing accessibility of AI tools has significantly lowered the barriers to academic dishonesty, presenting substantial challenges for educators and institutions striving to uphold rigorous ethical standards in student learning and assessment.
Because of these issues, Prather et al. \cite{prather2025beyond} propose reassessing educational methods and evaluation strategies. This involves developing teaching and assessment practices that reduce opportunities for academic misconduct while promoting genuine learning.
Eachempati et al. \cite{eachempati2025ShouldOE} suggest transitioning from written to oral examinations, proposing that oral exams should become the default assessment method.
Oral assessments offer a more direct way to verify student comprehension and understanding, while simultaneously fostering critical thinking and effective communication skills, abilities that are indispensable in the modern world.

\paragraph{Incomplete mental models}
Wang et al. \cite{wang2023exploring} suggest that the reliance on AI tools may contribute to the formation of incomplete mental models. This concern aligns with the concept of bricolage, as described by Ben-Ari \cite{ben2001constructivism}, where learning is a process of accumulating facts that progressively coalesce into a coherent model.
However, it is plausible that alternative learning approaches, particularly those heavily dependent on AI, could lead to the construction of divergent or even flawed models. 
As a potential solution, Hoffman et al. \cite{hoffman2023measures} advocate for the adoption of explainable AI. 
In the context of explainable AI, users are not merely presented with the outcomes; rather, they have access to the underlying steps and reasoning that lead to these results. This transparency allows for verification and comprehension of the AI's decision-making process, enabling users to critically evaluate the validity of the outcomes and the logic behind them.

\paragraph{Reduced class engagement}
Nie et al. \cite{nie2024gpt} observed that while the integration of AI tools led to improved learning outcomes, it was accompanied by a noticeable decline in classroom engagement. Similar findings were reported by Mollick \cite{ethan23} who noted that students often chose to direct their questions to AI systems rather than participating in peer or instructor-led discussions. The extent to which this reduction in engagement constitutes a significant problem remains unclear; it may be a transient side effect of introducing new technologies. Nonetheless, it raises concerns, as active participation and the development of interpersonal and communication skills are essential components of a well-rounded education and vital for functioning in society.

\paragraph{Shifting roles}

Bloom's taxonomy, initially devised for teaching and assessment, has the potential to offer a framework for facilitating critical thinking, as discussed by Gonsalves \cite{gonsalves2024GenerativeAI}.
However, Gonsalves \cite{gonsalves2024GenerativeAI} argues that the taxonomy’s linear, hierarchical structure may no longer align with the evolving collaborative dynamics between students and AI tools. In this shifting educational landscape, students are not passive recipients of information but active participants in the learning process, co-constructing knowledge alongside technological systems.
This shift from passive learning to active collaboration compels both educators and academic institutions to reconsider the traditional role of students, reevaluate curriculum designs, and rethink assessment methods.
This transformation presents significant challenges for academia as institutions must grapple with entrenched structures and pedagogical norms while balancing the integration of AI tools with the development of human-centric capabilities.
Achieving this necessitates a fundamental shift in how knowledge is constructed, assessed, and applied, calling for a reconceptualization of both the role of academia and that of educators.
Moreover, it necessitates reevaluating education's broader objectives: How can it prepare students to meet the complex and evolving demands of the labor market and contemporary society?

\subsection{Summary}

CS educators face many potential issues related to the adoption of AI in their teaching.
It is entirely possible that to address at least some of these problems, the way we organize teaching and learning has to change.
It is also likely that the skills we should teach will be different in the future to address how AI is altering the way CS students should approach problems during their careers.

\section{Discussion}
\label{sec:discussion}
The central question is not whether to incorporate AI into educational practices, but how to do so in a manner that preserves and enhances educational quality, ensuring that students not only adapt to new tools but achieve deeper, more meaningful learning outcomes that are relevant and applicable in the rapidly evolving labour market and society. Addressing this challenge requires more than the integration of novel technologies into existing pedagogical frameworks;  it calls for a critical reassessment of the underlying purposes, practices, and institutional logics that shape not only educational systems but also the broader societal structures within which they operate.

\subsection{What is the potential impact of AI on CSEd?}

Whether embraced or approached with caution, AI is already exerting a transformative impact not only in CSEd but also on the broader technological, economic, and epistemic fabric of society.
In CSEd, AI tools (from code generation assistants to agents and intelligent tutoring systems) are reshaping the ways students engage with programming, problem-solving, and computational thinking.


A central concern is whether students are acquiring the deeper conceptual and architectural understanding necessary for sustainable software development.
As tools like GitHub Copilot and ChatGPT automate syntax and low-level implementation details, there is a risk that learners may bypass the cognitive struggle essential to developing robust computational thinking skills.
Rather than writing every line of code, students now increasingly interact with systems that suggest, complete, or even design code on their behalf.
While this may streamline productivity, it also raises foundational questions about what it means to learn programming in an AI-mediated world.
Without deliberate instructional scaffolding, students may become overly reliant on AI-generated solutions and fail to develop critical skills in abstraction, systems design, and debugging, skills that are vital for constructing and maintaining complex software architectures.

Dismissing AI tools outright would be equally shortsighted. The challenge lies not in avoiding these technologies, but in reorienting curricula to enable critical engagement with them.
Students must learn not only how to use AI tools, but also when and why to use them, and with what potential implications.
Educational institutions play a crucial role in supporting this by cultivating students’ understanding of how AI systems function, including their training data, output generation methods, and susceptibility to failure or bias.

These developments signal broader ontological shifts in the labor market and computational work.
Future software engineers must possess not only technical proficiency but also adaptive expertise, the capacity to collaborate with AI systems, critically evaluate their outputs, and design high-level solutions that align with human values and long-term system objectives.
The demands placed on various educational roles are likely to shift in response to these developments, and in \autoref{table:ai-usages-in-he} we outline potential changes in responsibilities and practices across the higher education landscape.


\begin{table*}[!b] 
  \def\arraystretch{1.5}
  \caption{Potential impacts of AI on educational roles and corresponding strategic responses}
  \label{table:ai-usages-in-he}
  \begin{tabular}{p{1.5cm} p{2.9cm} p{4.3cm} p{7.8cm}}
    \toprule
    Role & Description & Usages & Vision \\
    \midrule
    Managerial & Managerial roles deal with the big picture and system management. & Learning analytics \cite{demartini2024artificial}, identifying learning patterns, curriculum and system design, student management and clustering, personality profiling \cite{crompton2023artificial}. & AI can provide the feedback needed for curriculum and course design while capturing student analytics to see how specific actions performed at the managerial level impact learning results, for example. AI can help generate new insights to allow the crafting of relevant and efficient learning programs. \\
    Instructive & Instructive roles deal specifically with teaching and relaying information to students. & Automated grading, self-regulated learning, personalized feedback, reviewing online activities of students, evaluating educational resources, and generating tests \cite{crompton2023artificial}. & AI-based approaches can give instructors more personalized ways to instruct their students. Most importantly, AI can help to gain insights into how the students are learning to allow adjustment of the course material to suit students better. \\
    Predictive & Predictive level lets us know what might happen in the future. & Predicting academic performance and dropout risk. & AI can extract information to detect patterns and predict the future. Predictions come with ethical concerns related to what should be analyzed and who should receive the predictions. \\
    Learning & Learning captures the student's perspective. & Personalized learning and support, as well as immediate assistance. & AI agents are already used in different contexts \cite{taneja2024jill, markel2023gpteach, zhu2021teaching} to support learning. A good agent could become a partial replacement for a teaching assistant and act as first-tier support. \\ 
  \bottomrule
\end{tabular}
\end{table*}

\subsection{Risks and challenges of using AI in CSEd}

Mollick and Mollick \cite{mollick2023using} emphasize the importance of integrating AI tools into teaching with careful consideration of their pedagogical value. They argue that the focus should remain on how such tools can meaningfully support learning, rather than on the novelty or capabilities of the technology itself. 
MacArthur \cite{macarthur2025large} frames this issue as a need to maintain a clear focus on developing students’ critical reading, writing, and editing skills, which remain essential for successfully engaging with AI tools. Without these foundational competencies, excessive reliance on AI could potentially undermine learning outcomes, a concern substantiated by preliminary findings from Kosmyna et al. \cite{kosmyna2025your}.
These findings highlight the importance of how AI tools are integrated into teaching; the chosen pedagogical approach plays a crucial role in shaping their educational effectiveness.

To facilitate better learning when integrating AI tools and to minimize negative cognitive loading, it is vital to adopt thoughtful approaches and select appropriate tools.
Here, careful curation and integration of AI tools are critical to improve student engagement and reduce cognitive overload.
Additionally, incorporating metacognitive strategies can help students become aware of their learning processes.
Teaching students to reflect on their interactions with AI tools allows them to identify areas for improvement and develop self-regulation skills.
Furthermore, fostering collaborative learning environments can facilitate peer interaction and knowledge sharing, creating opportunities for students to learn from one another and collaboratively address issues, thus reducing individual cognitive strain.
Lastly, it is essential to continually evaluate and adapt the integration of AI tools within the learning process.
Feedback from students, along with a systematic assessment of the impact of AI tools on learning outcomes, can help educators and institutions make informed decisions about how to refine the use of such technologies to better align with pedagogical goals.
A central challenge in this process lies in defining appropriate "metrics of success."


For educators introducing AI technologies into learning environments, it is not sufficient to simply acknowledge the existence of AI-related biases; rather, it is necessary to situate them within broader sociotechnical and epistemological frameworks.
AI systems are not neutral instruments — they are shaped by the data on which they are trained and the design choices embedded by their developers, often reproducing and amplifying existing social, cultural, and institutional inequalities. 
From this perspective, the classroom becomes a critical site for examining how these systems mediate knowledge production and power relations.
Educators must therefore engage students in a systematic interrogation of how biases manifest, how they differentially impact learners and society more broadly, and how they intersect with issues of access, representation, and agency.
This stance moves beyond a surface-level awareness of biases toward cultivating the critical literacies necessary to navigate, contest, and reshape AI’s role in education.

\subsection{How could CS curriculum change to take the impact of AI into account?}

Given AI poses new challenges for CSEd, likely the way we teach has to change as well.
The key point is to use methods that support learning while imparting key skills to our students that help them to leverage AI tools sensibly while understanding their caveats.
We have considered several complementary directions in \autoref{tab:cs_directions}.

\begin{table*}[!b] 
    \def\arraystretch{1.5}
    \caption{Addressing the impact of AI through alternative teaching and assessment methods}
    \label{tab:cs_directions}
    \centering
    \begin{tabular}{p{4.5cm} p{12.5cm}}
        \toprule
        Direction & Description \\
        \midrule
         Rethinking the role of the classroom & As noted by Wang et al. \cite{wang2023exploring}, rethinking the role of the classroom, by using, for example, flipped learning, may be beneficial as it puts focus on collaboration during contact time with students while allowing them to understand content at their own speed. The approach goes well with AI tutors that enable fast feedback loops, although educators have to take care to calibrate them well. \\
         Oral examinations and dialogical assessment & As mentioned by Eachempati et al. \cite{eachempati2025ShouldOE}, moving to oral examinations allows educators to probe understanding in real-time although they come with challenges related to scalability and accessibility. \\
         Socratic discussion seminars & Seminars build upon oral examinations and emphasize dialogical reasoning, critical questioning, debate, and peer-based exploration of complex topics. In CSEd, Socratic sessions can revolve around topics such as data ethics, the epistemology of algorithms, or debates on computational creativity. \\
         Team-based hackathons and collaborative design challenges & In hackathons, students collaboratively design and implement solutions to open-ended problems under time constraints while using AI to support the process. The challenge is figuring out how to assess students, and this direction should be combined with others to achieve that. \\
         Closed-network examinations with contextual constraints & By applying resource constraints, students may experience environments that mirror real-world limitations. The approach allows evaluating baseline understanding of students, as usage of AI may be disallowed temporarily, for example. \\
         Reflective computational journals & By documenting their learning, students gain a view of their learning process. The resulting journal can be used as part of the evaluation. As hinted by Mollick and Mollick \cite{mollick2023using} and Znamenskiy et al. \cite{znamenskiy2025integrating}, a part of this could be evaluating AI outputs critically and documenting the findings. \\
         Code comprehension and debugging challenges & Given AI tools make the creation of code cheap, shifting focus to code comprehension and debugging is a worthwhile direction since it relies on critical evaluation and problem solving, both important skills in working life. \\
         Design-driven capstone portfolios & Developing a portfolio of work during studies allows students to showcase their skills. The key point is to encourage students to create original work with thoughtful design through holistic evaluation to discourage overuse of AI solutions to set them apart. \\
         \bottomrule
    \end{tabular}
\end{table*}

\subsection{What might the future of CSEd look like?}

Inayatullah's Futures Triangle \cite{inayatullah2023futures} can provide a useful framework for scenario building by analyzing three key forces: the pull of the future, the push of the present, and the weight of the past. 
This model could serve as a valuable conceptual tool for understanding the dynamics shaping CSEd, allowing reflection on current developments and offering insight into how these dynamics may influence its future trajectory.

\begin{description}
  \item[Weight of the past] \hfill \\ Academic tradition, resistance to adopting new technologies, legal framework, and ethical issues
  \item[Push of the present] \hfill \\ Adoption of AI technology by students/universities/ICT industry, breakthroughs in technology, and shifting technosocial landscape
  \item[Pull of the future] \hfill \\ Fully individualized learning in mass scale, low cost education for all, co-intelligence \cite{mollick2024co} or collective intelligence \cite{macarthur2025large} as the new normal, and reframing of human work
\end{description}


Even with a conservative outlook, it is clear that CSEd is on the brink of significant transformation, driven by the ongoing integration of AI technologies, and we can expect comprehensive reforms in CSEd programs designed to incorporate AI tools and methodologies.

However, this transition will not be without significant challenges. Both the AI technology itself and the teaching practices we use are evolving at a rapid pace, but various motivations are often misaligned across different sectors. The disconnect between commercial interests, academic objectives, and societal needs creates a complex web of challenges. Effective alignment will require not only thoughtful integration of technology into the curriculum but also extensive dialogue across political, institutional, and societal levels. 

Engaging with the future of CSEd remains an essential and strategic undertaking. By examining emerging trends, technological advancements, and shifts in pedagogy, these reflections can contribute to the development of frameworks that shape how CSEd will adapt to the evolving needs of both the workforce and society. These discussions are not merely speculative; they provide the foundation for evidence-based decision-making that will guide institutional strategies, curriculum design, and the thoughtful integration of emerging tools into the educational landscape.

\section{Conclusion}
\label{sec:conclusion}
To explore the transformative potential of AI in CSEd, we addressed two research questions: one focused on prevailing perspectives in the field, and the other on potential challenges. The following sections present our findings, and finally, we identify key open research questions to explore potential research directions for the future.

\subsection{RQ1: What are the prevalent perspectives surrounding the adoption of AI within CSEd?}

In the context of adopting AI within CSEd, the community expresses both AI-critical and AI-positive perspectives.
AI-critical perspectives highlight several potential risks associated with the integration of AI tools. 
These concerns include the misuse of technology, challenges to effective learning practices, pedagogical reductionism, potential cognitive impacts, and the dangers of uncritical adoption.
AI-positive perspectives acknowledge that artificial intelligence is a permanent part of the educational landscape and enables new types of teaching approaches that were beyond our reach earlier.
Proponents argue that educators have a responsibility to harness their potential to enhance teaching and learning while thoughtfully addressing any associated challenges.




Artificial intelligence holds considerable promise for enhancing equitable access to quality education, particularly in under-resourced regions.  As AI technologies become increasingly cost-effective, they may offer scalable, adaptable solutions for addressing persistent gaps in educational provision, especially in contexts where access to qualified educators and adequate infrastructure remains limited. While human teachers will always be central to the learning process, AI can serve as a complementary resource and force multiplier, especially as both hardware and software become increasingly accessible. In this way, AI may help address disparities in access to quality education without replacing the essential social and pedagogical functions of educators.

In sum, the AI-positive perspective envisions a transformative role for AI in CSEd, grounded in the ability to personalize learning experiences, scale instructional support, and contribute to more equitable global access to high-quality education.
AI-critical perspectives acknowledge the risks related to the technology and keep the discussion about the adoption of AI grounded.

\subsection{RQ2: What are the issues potentially hindering adoption of AI in CSEd?}

We identified several factors that may hinder the adoption of AI in CSEd, with additional challenges likely remaining underexplored. Prominent concerns raised in the literature include increasing inequality, technological limitations, algorithmic biases, and learning-related difficulties. Among these, it is essential to prioritize the most critical issues, particularly as learning-related challenges have the potential to undermine the very benefits AI is intended to bring to education. 

Furthermore, the integration of AI tools in CSEd invites a re-examination of the roles of both educators and students, as well as the dynamics between them. Interacting with AI is inherently complementary to human intelligence, requiring a nuanced understanding of its capabilities and limitations. Realizing AI’s full potential in CSEd depends not only on effective tool usage but also on preserving and cultivating essential human skills that remain central to meaningful learning. To maximize the benefits of AI while mitigating its drawbacks, it is important to reconsider established pedagogical practices and remain open to alternative modes of instruction. This may include revisiting traditional approaches such as the Socratic method and oral examinations, methods that emphasize critical thinking, dialogue, and real-time reasoning. These approaches should not be dismissed as outdated but rather seen as vital components of a modern pedagogical toolkit, helping to create effective teaching strategies that prepare students for the challenges of an increasingly technology-driven world.







\subsection{Open research questions}

Although this paper gave insight into the potential and pitfalls of AI in CSEd, it also left many important questions open:

\begin{enumerate}
    \item What core skills should students acquire to make the most of AI, and what are the best methods for teaching these skills?
    \item How should CSEd be structured to fully leverage recent developments? 
    \item What are the key points of intersection between CSEd and AI across various levels of the education system?
    \item What national or cultural differences exist in the adoption of AI in CSEd, and how do these differences manifest in practice?
    \item How can we integrate AI into CSEd in a way that prevents widening inequalities and ensures that disadvantaged students benefit from these advancements as well?
\end{enumerate}

In addition to these questions, it would be valuable to adopt a longitudinal perspective on the topic to assess the long-term effects of integrating AI into CSEd.
Such an approach would allow us to better understand how AI influences student outcomes, teaching practices, and the broader educational landscape over time.

\subsection{Acknowledgments}

Special thanks to Tapio Pitkäranta, Carl Kugblenu, Pekka Enberg, Alimujiang Abudumiti, Misa Bakajic, and Astrid Schneider for feedback and useful pointers.

\printbibliography

\end{document}